\documentclass[prb,  onecolumn, square, aps] {revtex4}
\usepackage[dvips]{graphicx}

\begin{document}

\title{Metal-to-glass ratio and the Magneto-Impedance of
glass-covered CoFeBSi microwires at high frequencies}

\author{R. Valenzuela\P, A. Fessant, J. Gieraltowski and C. Tannous} 
 \affiliation{Laboratoire de 
Magn\'etisme de Bretagne, CNRS-FRE 2697, Universit\'e de Bretagne Occidentale, BP: 809 Brest CEDEX, 29285 France, \\
\P Materials Research Institute, National University of Mexico, P.O. Box 70-360, 
Coyoacan, 04510 Mexico}

\begin{abstract}
High frequency [1-500 MHz] measurements of the Magneto-Impedance (MI) of 
glass-covered Co$_{69.4}$Fe$_{3.7}$B$_{15.9}$Si$_{11}$
microwires are carried out with various metal-to-wire diameter ratios. 
A twin-peak, anhysteretic behaviour is observed as a function of magnetic 
field. A maximum in $\Delta Z/Z$ appears at  different values of the frequency $f$,
125, 140 and 85 MHz with the corresponding diameter ratio $p$ = 0.80, 0.55 and 0.32. 
We describe the measurement technique and interpret our results with a thermodynamic
 model that leads to a clearer view of the effects of $p$ on the maximum value of MI and the 
anisotropy field. 
\end{abstract}


\maketitle

\section{Introduction}

When a ferromagnetic conductor is traversed by a current of low amplitude 
and high frequency, its impedance or rather  its Magnetoimpedance (MI) 
can be altered by applying a dc magnetic field. \\
This phenomenon, first described \cite{Harrison} in the 1930's, has been receiving special 
attention over the last 15 years \cite{Makhotkin,Mandal} due to its potential technological 
applications \cite{Mohri,Vazquez} in sensors, devices and instruments. Its fundamental physics
is also being deeply examined  \cite{Knobel}. \\

MI has been observed in a wide variety of materials, geometries and structures,
particularly in amorphous wires having typically diameters of a few
hundred microns. Wires with smaller diameters (a few microns)
covered with a glass sheath show an increase of the working 
frequency, and introduce an additional structural feature that alter
the physical parameters. Since glass exerts some mechanical stress on the 
metallic wire, a change in the magnetic response is expected. Therefore it is
of interest to finely tune the physical properties through the control of the
thickness and nature of the glass sheath. \\

In this paper, we report on MI measurements of Co-rich amorphous 
microwires with various ratios of the metal-to-glass diameter, in the 
[1-500 MHz] frequency range, carried out with a novel \cite{Fessant} broadband 
technique. This technique allows a complete determination of MI as a 
function of both frequency and magnetic field. The effects of the thickness
of the glass sheath  are clearly illustrated and the variation of the anisotropy
field $H_K$ is evaluated directly as a function of stress.

\section{Experimental results}

Glass-covered amorphous microwires of nominal composition 
Co$_{69.4}$Fe$_{3.7}$B$_{15.9}$Si$_{11}$ were prepared by fast cooling with 
the glass-coated melt-spinning technique also known as Taylor-Ulitovski technique. 
Several metal-to-wire ratio values, $p=\phi_m/\phi_w$, 
with $\phi_m$ the metallic core diameter and $\phi_w$ the total wire diameter,
were produced and characterized.
For values of metal core diameters of 24, 12 and 7 $\mu $m, 
with corresponding  total diameters of 30, 21.8 and 21.9 $\mu $m we get
$p$ = 0.80, 0.55 and 0.32, respectively. In order to 
make electrical contacts, the glass sheath was etched away over a few mm on both 
microwire ends, with a solution of hydrochloric acid. Silver paste contacts were then
made in order to proceed with the electrical measurements. \\

MI measurements were carried out in the [1-500 MHz] range, on pieces of 
microwires $\sim $12 mm in length, with an HP 8753C Network Analyzer using a novel 
broadband measurement technique described in \cite{Fessant}. Helmholtz coils served as
source of axial dc  magnetic fields ranging from -80 Oe to 80 Oe. \\

The results obtained are typically  plotted in a 3D representation of $\Delta Z/Z$:
\begin{equation}
\Delta Z/Z = (Z_{H_{dc}= 0 \hspace{0.5mm}Oe}-Z_{H_{dc}= 80 \hspace{0.5mm}Oe}) / 
Z_{H_{dc}= 80\hspace{0.5mm}Oe}
\end{equation}

\begin{figure}[!h]
\begin{center}
\scalebox{0.3}{\includegraphics[angle=0]{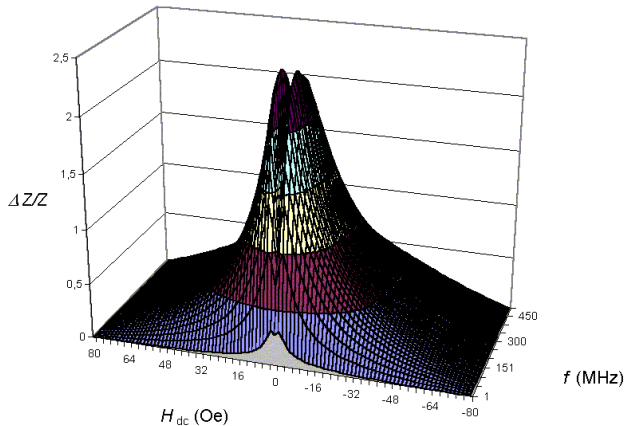}}
\end{center}
\caption{Magnetoimpedance plot for $p$ = 0.8, as a function of axial dc field 
and frequency. The typical double-peak MI plot is observed, with no hysteresis under 
inversion of $H_{dc}$.}  
\label{fig1}
\end{figure}

where $Z$ is the total impedance modulus [$Z = \sqrt{(Z'^{2}+Z''^{2})}$ ; with $Z'$ 
the real part and $Z''$ the imaginary part of impedance], as a function of 
dc field, $H_{dc}$, and frequency, $f$. The results are shown for $p$ = 0.8 in 
Fig. 1. The expected symmetrical double peak MI plot is obtained as a function of $H_{dc}$; 
the peaks are associated with $\pm H_K$, the anisotropy field. We obtain $H_{K}
\sim  $ 3.5 Oe and no hysteresis was observed by cycling the dc field $H_{dc}$. Regarding 
frequency $f$, MI shows a maximum of $\sim $ 250 {\%} at about 100 MHz. 
Similar plots were obtained with the other $p$ ratios, albeit with significant 
differences in the values of the anisotropy field and peak frequency values. 
This allows us to make a detailed comparison, for a fixed frequency (as is typically presented),
of the effect of the magnetic field. For instance, Fig. 2. displays the results we 
obtain at 10 MHz.  The diameter ratio seems to produce a strong damping effect
on the MI response. \\

\begin{figure}[!h]
\begin{center}
\scalebox{0.3}{\includegraphics[angle=0]{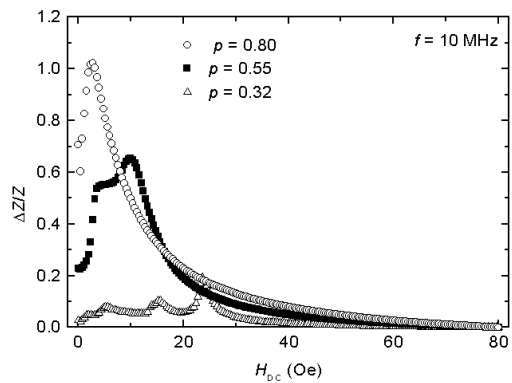}}
\end{center}
  \caption{MI as a function of $H_{dc}$, at a fixed frequency of 10 MHz. Note the 
  broad distribution of the value of $H_K$ for small values of $p$.}  
\label{fig2}
\end{figure}

\begin{figure}[!h]
\begin{center}
\scalebox{0.3}{\includegraphics[angle=0]{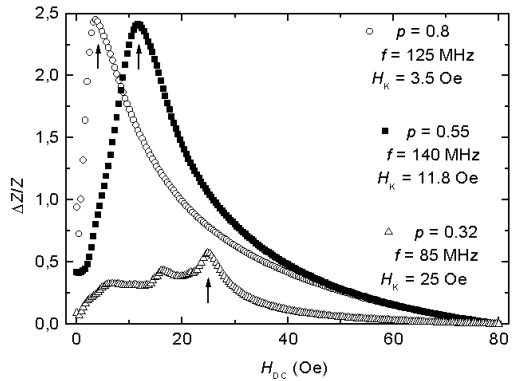}}
\end{center}
\caption{$\Delta Z/Z$ plot as a function of the applied dc field $H_{dc}$ at
the selected frequency of the MI $\Delta Z/Z$ maximum, for each diameter ratio $p$.
Note the broad distribution of the value of $H_K$ at small values of $p$.}  
\label{fig3}
\end{figure}

Since our measurement method provides information over the full [1-500 MHz]
range, we can use a deeper physical basis to make such a comparison. We 
choose the frequency at which the maximum in $\Delta Z/Z$ appears and compare 
results as a function of $H_{dc}$, as shown in Fig. 3. In addition to the 
absolute differences in MI values, important changes in the relative values 
are observed, as compared with Fig. 2. Now the $p$ = 0.55 microwire shows a MI 
maximum close to that of $p$ = 0.80. The anisotropy field, however, is 
three times larger. Note that the sample with $p$ = 0.32 exhibits, as a 
function of field, several peaks that can be associated with a distribution of the 
anisotropy axis orientation. This introduces a large uncertainty in the numerical 
value of the anisotropy field $H_K$ that will be discussed further below. \\

The effect of $p$ on the MI response and anisotropy field is roughly indicated in Fig. 4. 
This is consistent with what has been previously observed: MI response increases 
as $p$ increases, while the anisotropy field decreases. $p$ indicates the importance
of the metal core with respect to the total diameter of the wire and the stress
increases as the thickness of the glass sheath increases. \\
The result is consistent with the following fact.
During fabrication, glass-covered microwires are subjected to strong 
stresses, generally proportional to the thickness of the glass coating that varies
inversely proportional to $p$. The origin of such stresses can be 
somehow, readily understood, since glass possesses a smaller thermal contraction coefficient 
than metals. In the cooling process, the metallic core tends to contract faster and 
more substantially  than the surrounding glass sheath, however glass hampers 
such contraction. The overall consequences and the nature of the stresses are highlighted
in the next section.

\section{Interpretation of the results and Conclusions}
Torcunov et al. \cite{Torcunov} modeled the thermoelastic and quenching stresses that occur
in glass-coated wires and evaluated with a thermodynamic model the stress components in terms of their 
axial $\sigma_{zz}$, radial $\sigma_{rr}$ and azimuthal  $\sigma_{\phi \phi}$ components 
(in a cylindrical system of coordinates $(r,\phi,z)$ with the $z$ axis along the wire).
The following expressions (providing the Poisson's coefficients of the glass and metal are
equivalent $\nu_g \sim \nu_m \sim \frac{1}{3}$) are obtained and adapted to our case:
\begin{eqnarray}
\sigma_{rr} & = & \frac{\epsilon E_{g} (1-p^2)}{(\frac{k}{3} +1) (1-p^2) +\frac{4 p^2}{3} } \\
 \sigma_{\phi \phi} & = &  \sigma_{rr}  \\
 \sigma_{z z} & = &  \sigma_{rr}  \frac{(k+1)(1-p^2) + 2 p^2}{k (1-p^2) + p^2} \\
\end{eqnarray}

where $E_{g}$ is the glass Young modulus, $k=E_{g}/E_{m}$ and $E_{m}$ is the metallic wire
Young modulus. \\

The term $\epsilon$ is given by the difference of the glass
and metal expansion coefficients $\alpha_g, \alpha_m$ (respectively) times the difference
of the minimum glass solidification  temperature $T^{*}$ and room temperature $T$,
$\epsilon=(\alpha_m-\alpha_g)(T^{*}-T)$. \\

\begin{figure}[!h]
\begin{center}
\scalebox{0.2}{\includegraphics[angle=0]{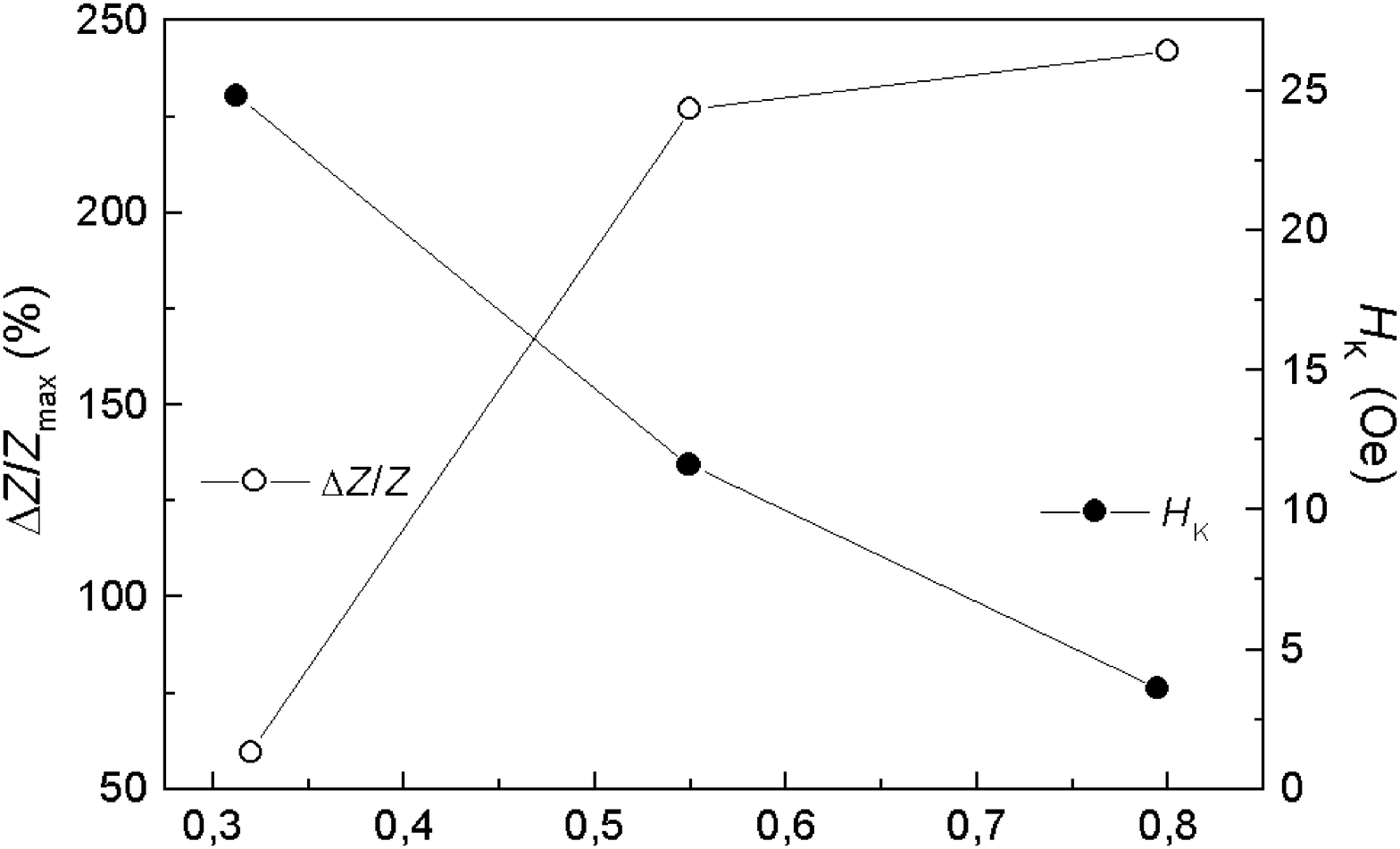}}
\end{center}
\caption{Overall indication of the effect of diameter ratio, $p$, on the MI maximum value 
and on the anisotropy field $H_K$. In the following figures, we show that the $H_K$ value
is overestimated at small $p$.}  
\label{fig4}
\end{figure}

\begin{figure}[!h]
\begin{center}
\scalebox{0.5}{\includegraphics[angle=0]{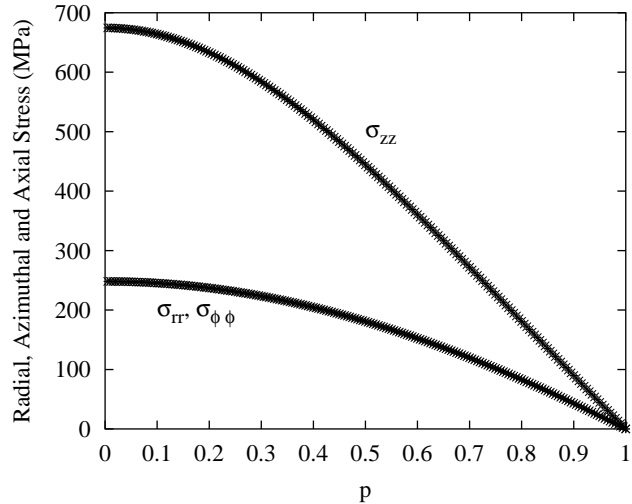}}
\end{center}
\caption{Effect of diameter ratio, $p$, on the axial $\sigma_{zz}$, radial $\sigma_{rr}$ and 
azimuthal  $\sigma_{\phi \phi}$ stress components. Parameters used are taken from Adenot et al. \cite{Adenot}.
All stress components decrease with $p$ as predicted.}  
\label{fig5}
\end{figure}

\begin{figure}[!h]
\begin{center}
\scalebox{0.5}{\includegraphics[angle=0]{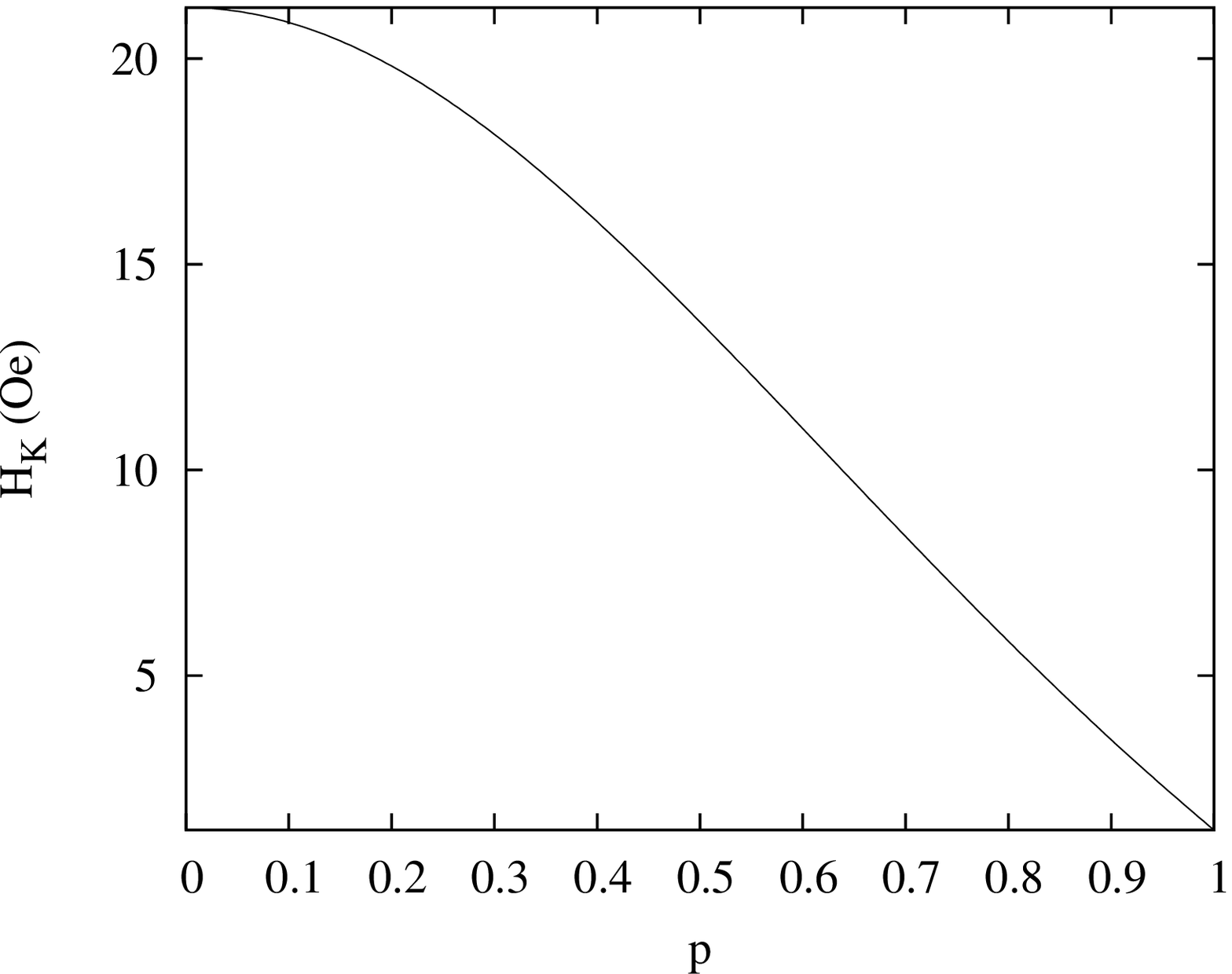}}
\end{center}
\caption{Effect of diameter ratio, $p$, on the anisotropy field $H_K$. The parameters taken from
Mohri et al. \cite{Humphrey} are $\mu_0 M_s$=0.8 T, zero stress anisotropy constant $K_{\sigma=0}$ =40 J/m$^3$ and 
$\lambda_s$=-0.1 10$^{-6}$. All the other parameters are the same as in Figure 5. As $p$
decreases, the tapering off of $H_K$ is faster than expected, nevertheless an  increase in
experimental uncertainty should be accounted for as well. For large values of $p$ good agreement
between theory and experiment is obtained.}  
\label{fig6}
\end{figure}

Using the numerical values \cite{Humphrey}, 3.2 10$^{-6}$  K$^{-1}$ and  1.2 10$^{-5}$  K$^{-1}$ for 
the expansion coefficients $\alpha_g, \alpha_m$, $T^{*}=550$C, 64 GPa  and 110 GPa
for the Young modulus of glass and metal $E_{g}, E_{m}$ we get the variation 
of all stress components versus $p$ as depicted in Fig.5. The variation shows that
all stresses decrease steadily with the increase of $p$ as one might expect, since
the origin of stress is associated with the increase of glass thickness. \\
 
Applying this variation to the anisotropy field $H_K=2K_\sigma/\mu_0 M_s$ with $K_\sigma$ the
anisotropy constant of the wire under stress $\sigma$ and adopting the change
of the anisotropy constant according to \cite{Adenot} with the additional assumption of no extra
applied stress ($\mu_0$ and  $M_s$ are vacuum permeability and saturation magnetization
respectively):

\begin{equation}
K_\sigma= K_{(\sigma=0)} -\frac{3}{2}\lambda_s (\sigma_{zz} - \sigma_{\phi \phi}) 
\end{equation}  

Using physical parameters of wires \cite{Humphrey} with a composition
(Co$_{0.94}$ Fe$_{0.06}$)$_{72.5}$ B$_{15}$ Si$_{12.5}$ similar to ours,
we get in Fig.6, a reasonable agreement with the experimental behaviour 
depicted in Fig.4, despite a faster tapering off of $H_K$ at low values of $p$
where we observe experimentally a large uncertainty in the value of  $H_K$ due
to a broad distribution of anisotropy axis orientation (see Figs.2 and 3). From
Fig.6, we infer that when $p=0.32$, the value of $H_K$ has been overestimated and
should be about 20 Oe and not 25 Oe as obtained in Figs.2 and 3.   \\

In conclusion, the measurement of the MI response of microwires with a novel 
broadband technique provides a satisfactory view of the interplay between
different physical phenomena operating on the glass or metal side. It is observed
that when the metal core is small ($p$ small) a larger distribution of $H_K$ is 
observed. One possible cure to that problem is to perform the Taylor-Ulitovsky 
process under the application of a magnetic field in order to control magnetic 
orientation during growth or to perform post-annealing with/out external stress.
Several applications of the present results are possible. One of them is
the ability to select or tune the physical properties such as a better microwire 
might be produced and  suited  for a specific application. A simple description
of the desired property might be  given by a
$\Delta Z/Z$ percentage value, static field or frequency operation range and finally
a sensitivity, bandwidth or signal to noise figure of merit. \\

{\bf Acknowledgements} \\
The authors acknowledge Prof. M. Vazquez (Spain) for providing the 
microwires samples; R.V. thanks DGAPA-UNAM, Mexico, for partial support 
through grant PAPIIT IN119603-3.

\end{document}